\documentclass[12pt]{article}
\usepackage[pdftex]{graphicx}
\usepackage{amsmath}
\usepackage{amssymb}
\usepackage{euscript}
\usepackage{multirow}
\usepackage{color}
\usepackage{colortbl}
\usepackage{calc}
\usepackage{cite}

\newcommand{\figtxt}[1]{\footnotesize{#1}}

\makeatletter
\newlength\twolinebox@linelength
\newlength\twolinebox@columnheight
\newcommand{\twolinebox}[2]{%
   \setlength{\twolinebox@linelength}%
             {\maxof{\widthof{#1}}{\widthof{#2}}}%
   \setlength{\twolinebox@columnheight}{\heightof{#1}+\depthof{#1}+0.2em+0.4em/2+\heightof{0}/2}%
   \raisebox{0pt}[\twolinebox@columnheight][\heightof{\vbox{\vskip0.2em\hbox to 
   \twolinebox@linelength {#1\hfil}\vskip0.4em\hbox to 
   \twolinebox@linelength {#2\hfil}}}+\depthof{\vbox{\vskip0.2em\hbox to 
   \twolinebox@linelength {#1\hfil}\vskip0.4em\hbox to 
   \twolinebox@linelength {#2\hfil}}}-\twolinebox@columnheight+0.2em]{\vbox to 
   \twolinebox@columnheight{\vskip0.2em\hbox to 
   \twolinebox@linelength {#1\hfil}\vskip0.4em\hbox to 
   \twolinebox@linelength {#2\hfil}}}%
}

\newcommand{\tw}      {\textwidth}

\newcommand\WINHAC[0] {\textsf{WINHAC}}

\newcommand\PYTHIA[0] {\textsf{PYTHIA}}
\newcommand\HERWIG[0] {\textsf{HERWIG}}
\newcommand\HERWIGpp[0] {\textsf{HERWIG++}}
\newcommand\MC@NLO[0] {\textsf{MC@NLO}}
\newcommand\POWHEG[0] {\textsf{POWHEG}}

\newcommand\LHAPDF[0] {LHAPDF}

\newcommand{\flatDfDx}[2] {d\,{#1}/d\,{#2}}

\newcommand{\plus}   {{+}}
\newcommand{\minus}  {{-}}

\newcommand{\Wp}     {{W^+}}

\newcommand{\Wm}     {{W^-}}   

\newcommand{\etal}   {{\eta_l}}

\newcommand{\thetal}       {{\theta_l}}

\newcommand{\Asym}[1]   {\mathrm{Asym}^{(+,-)}\left(#1\right)}

\def\slashii#1{\setbox0=\hbox{$#1$}            
  \dimen0=\wd0                                 
  \setbox1=\hbox{\sl/} \dimen1=\wd1            
  \ifdim\dimen0>\dimen1                        
     \rlap{\hbox to \dimen0{\hfil\sl/\hfil}}   
     #1                                        
  \else                                        
     \rlap{\hbox to \dimen1{\hfil$#1$\hfil}}   
     \hbox{\sl/}                               
  \fi}

\definecolor{rltbrightred}{rgb}{1,0,0}
\definecolor{rltred}{rgb}{0.75,0,0}
\definecolor{rltdarkred}{rgb}{0.5,0,0}
\definecolor{rltbrightgreen}{rgb}{0,0.75,0}
\definecolor{rltgreen}{rgb}{0,0.5,0}
\definecolor{rltdarkgreen}{rgb}{0,0,0.25}
\definecolor{rltbrightblue}{rgb}{0,0,1}
\definecolor{rltblue}{rgb}{0,0,0.75}
\definecolor{rltdarkblue}{rgb}{0,0,0.5}
\definecolor{webred}{rgb}{0.5,.25,0}
\definecolor{webblue}{rgb}{0,0,0.75}
\definecolor{webgreen}{rgb}{0,0.5,0}
\definecolor{Black}{rgb}{0,0,0}
\definecolor{Greymax}{rgb}{0.65,0.65,0.65}
\definecolor{Greycen}{rgb}{0.75,0.75,0.75}
\definecolor{Greymin}{rgb}{0.85,0.85,0.85}
\definecolor{hl}{rgb}{
                0.909803922,       
                0.82745098,               
                0.909803922}

\begin{document}

\begin{titlepage}

~\vspace{1mm}

\begin{center}
{\LARGE\bf Charge asymmetries of lepton transverse momenta
in Drell--Yan processes 
at the LHC}
\end{center}

\vspace{4mm}

\begin{center}
{\large\bf  
 M.~W.\ Krasny$^{a}$ 
 {\rm and} 
 W.\ P\l{}aczek$^{b}$}

\vspace{4mm}
{\em $^a$Laboratoire de Physique Nucl\'eaire et des Hautes \'Energies, \\
          Universit\'e Pierre et Marie Curie Paris 6, Universit\'e Paris Diderot Paris 7, \\
          CNRS--IN2P3, \\
          4 place Jussieu, 75252 Paris Cedex 05, France.}
\\  \vspace{2mm}
{\em $^b$Marian Smoluchowski Institute of Physics, Jagiellonian University,\\
         ul.\ Reymonta 4, 30-059 Krakow, Poland.}
\end{center}

\vspace{10mm}
\begin{abstract}
Charged lepton transverse momenta in the Drell--Yan processes play an important
role at the LHC in precision measurements of the Standard Model parameters,
such as the $W$-boson mass and width, their charge asymmetries and  
$\sin^2\theta_W$. 
Therefore, their distributions should be described
as accurate as possible by the Monte Carlo event generators. 
In this paper we discuss the problem of matching the
hard-process kinematics of the Monte Carlo generator \WINHAC\ 
with the parton-shower kinematics of the \PYTHIA~{\sf 6.4} generator
while interfacing these two programs. 
We show that improper assignment  of the quark and antiquark  effective momenta 
in the LO matrix element computations may affect considerably the predicted lepton 
transverse momenta and even completely reverse their charge asymmetries
at the LHC. We propose two matching schemes in which 
the NLO QCD distributions of the leptonic kinematical variables  can be well reproduced
by the LO \WINHAC\  generator.      
\end{abstract}

\vspace{5mm}
\begin{center}
{\it To be submitted to Acta Physica Polonica B}
\end{center}

\vspace{18mm}
\footnoterule
\noindent
{\footnotesize
$^{\star}$The work is partly supported by the program of the French--Polish 
co-operation between IN2P3 and COPIN no.\ 05-116, 
and by the Polish National Centre of Science grant no.\ DEC-2011/03/B/ST2/00220.
}

\end{titlepage}

\section{Introduction}
\label{introduction}

Distributions of transverse momenta of charged leptons ($p_T^l$) produced
in Drell--Yan (DY) processes are important observables in hadron collider 
experiments. 
Their sensitivity to the values of the  Standard
Model (SM) parameters and to the polarisation of $W$ and $Z$ bosons produced in the DY process 
can be used  in precision tests of the Standard
Model (SM) and in searches for new physics. 

The measurement of the SM parameters: 
$W$-boson mass ($M_W$), its charge asymmetry ($M_{W^+} - M_{W^-}$) 
and width ($\Gamma_W$) at the LHC was investigated in 
Refs.~\cite{Krasny:2010vd,Krasny:2010zz,Fayette:2008wt,Fayette:2009pm,Fayette:2009zi,Rejzner:2009zz}. It was shown that in order to reach a precision
at the level of $\sim 10\,$MeV, the $p_T^l$ distributions
must be controlled experimentally to a 
comparable
accuracy. 
Since the Monte Carlo (MC) event generators are indispensable tools to derive the values of the SM parameters from the 
measured distributions, they must predict
these distributions to even higher precision. 

MC event generators for Drell--Yan processes developed so far can
be divided into two categories. 
The first one includes generators which are based on the 
precise calculations of matrix elements (ME)
for hard processes: those including  QCD effects to the NLO or even
NNLO level, and those including QED and electroweak (EW) corrections. 
In these generators, called hereafter the ME generators,  
differential cross sections for the hard process are convoluted with 
the universal parton distribution functions (PDFs) of the hadron beams. 
These PDFs depend, apart from the factorisation scale, only on 
the longitudinal momenta of partons, $x$.
In this type of the MC generators 
partons entering the hard process are assumed to be parallel to hadron
beam direction.
The second category  of MC generators includes the so-called parton-shower (PS)
generators, such as \PYTHIA\ \cite{Sjostrand:2006za}, 
\HERWIG\ \cite{Corcella:2000bw}, etc. 
They generate initial-state multi-parton radiation in form of the
LO-type QCD/QED parton cascade and then perform hadronisation as well as
some particle decays. In the PS generators, partons entering the hard process 
are no longer collinear with the hadron-beam directions, but acquire
non-zero transverse momenta. The hard process itself is described by these
generators usually at the LO level. Thus, as long as the ME  precision 
is the key factor determining the overall accuracy of the measurement of a  selected observable, 
they are inferior 
with respect to
the ME generators. However, in the remaining cases they often provide better 
description of the hadronic energy flow associated with the DY process, in particular
for not to high transverse momenta of the $W$ and $Z$ bosons. 

A very important, and at the same time difficult  issue is how to combine 
these two types of generators, avoiding, on the one hand, 
a double counting of QCD corrections and, on the other, possible gaps 
in phase space present in some PS algorithms. As the state-of-the-art
practical solutions to this problem for the QCD effects are regarded 
the \MC@NLO\ \cite{Frixione:2002ik} and \POWHEG\ \cite{Nason:2004rx} 
generators. They match the NLO QCD ME calculations
with the parton shower generators, albeit in different ways.  \MC@NLO\
uses \HERWIG\ or \HERWIGpp\ \cite{Bahr:2008pv} for parton shower generation, 
while \POWHEG\ is more universal, in principle it can use an arbitrary 
parton-shower generator.

In the case of ME generators that include QED/EW radiative corrections
the situation is simpler as combining them with PS generators usually does
not lead to double counting%
\footnote{The only problem here may be the QED ISR, but since its numerical
effects are rather small, it can be dealt sufficiently well by PS generators.}.
One of such generators is the MC program 
\WINHAC~\cite{WINHAC:MC,Placzek2003zg,Placzek:2009jy}. It includes higher-order
QED effects for the final-state radiation (FSR) and initial-final state
interferences in the Yennie--Frautschi--Suura (YFS)
exclusive exponentiation scheme \cite{yfs:1961} 
 together with the ${\cal O}(\alpha)$ EW 
corrections for the full charged-current DY process \cite{Bardin:2008fn}. 
For the QCD (and QED ISR) effects it is interfaced with
\PYTHIA\ {\sf 6.4} \cite{Sjostrand:2006za}. 
However, we do not use the Les Houches Accord (LHA) scheme \cite{Alwall:2006yp}
but our own interface, in which the \PYTHIA\ routines for parton-shower 
generation and hadronisation are called directly from the \WINHAC\ program.
The principal reason is that the LHA is not general enough 
to transmit the full information contained in the spin density matrix of the $W$ and $Z$ bosons 
between these two generators.
Moreover, from  the purely technical perspective, 
we avoid writing/reading events into/from disk
files, which makes event generation easier and much more efficient. 
For example, in
our studies presented in Refs.\cite{Krasny:2010vd,Krasny:2010zz,Fayette:2008wt,Fayette:2009pm,Fayette:2009zi,Rejzner:2009zz} and  requiring generation of 
${\cal O}(10^{11})$ events,  its efficiency was one of the principal optimisation targets. 

In any interface which extends the LHA scheme to processes involving  the spin-1 EW bosons as intermediate particles
a  particular  care must be taken of the  proper matching of the 
ME-type kinematics with the PS-type kinematics.
In the following we show that this is particularly important to describe the charged 
lepton transverse momenta distributions in the Drell--Yan processes at the LHC.
In particular, we find that using the original \PYTHIA\ effective momenta
of incoming quarks in the above matching results in strongly biased $p_T^l$ 
distributions, which is particularly visible in their charge asymmetries.
We propose solutions to this problem that seem to reproduce well the NLO QCD 
predictions for these asymmetries, as obtained e.g. from \MC@NLO. 
The goal of this exercise is to try to achieve the NLO QCD precision for  the description 
of the leptonic observables in the DY process with the suitably matched LO QCD generator 
which incorporates the state-of-the-art EW corrections.   

The paper is organised as follows. In the next section we describe in detail
the matching of the ME kinematics with PS kinematics as it is realised in the
\WINHAC\ MC event generator.
In Section~3 we present numerical results illustrating the above issues,
discuss their meaning for the LHC physics and propose our solutions
that match the NLO QCD predictions. 
Finally, in Section~4 we summarise the paper. 

\section{Matching of ME and PS kinematics}
\label{sec:matching}

The \WINHAC~\cite{WINHAC:MC} MC event generator is dedicated to precision
modelling of single $W$-boson production with leptonic decays, i.e.\ 
the charged-current Drell--Yan processes, in proton--proton,
proton--antiproton and ion--ion collisions, with the main emphasis on the
QED effects and electroweak corrections. It uses fully massive spin amplitudes
to evaluate the hard process matrix elements. They can be computed in
an arbitrary reference frame, in particular they can be used to calculate
polarised $W$-boson cross sections%
\footnote{\WINHAC\ provides several options for computing polarised $W$-boson
 processes in several  reference frames.}. 
In terms of the perturbative QCD the current version includes the LO hard-process 
matrix  element. The QCD effects enter only through 
scaling-violating PDFs, taken from the \LHAPDF\ library 
\cite{Whalley:2005nh}. Therefore, in \WINHAC\   incoming 
quarks producing a $W$-boson are collinear with hadron (ion) beams;
their longitudinal momenta are given by the two $x$-variables which are
generated according two PDFs and subsequently convoluted with the hard-process differential
cross section. At this stage \WINHAC\ is a ME-type MC generator.
Its full event kinematics (i.e.\ all four-momenta of initial, intermediate and 
final state particles) is constructed for incoming partons collinear with
the colliding beams. Let us call it the {\em ME kinematics}.
At this level \WINHAC\ has been cross-checked numerically to a high precision
with independent calculations 
\cite{Placzek2003zg,CarloniCalame:2004qw,Bardin:2008fn}.

Events of the {\em ME kinematics} do not look very realistic from the 
experimental point of view
for the following two reasons. Firstly, QCD radiation affect not only longitudinal
momenta of partons but also their transverse momenta.
Therefore, using purely collinear PDFs for the description of the QCD effects is
not sufficient. Secondly, partons are not observed experimentally. What can be  observed 
are the  products of their hadronisation and decays. 
Therefore, in a realistic
MC generator, to be used in an experimental data analysis, the above  effects
must be taken into account. In \WINHAC\ this is done through the
interface to the \PYTHIA~{\sf 6.4} generator which performs the initial-state 
LO-type QCD (and QED) parton shower, appropriate proton-remnant treatment,  and
necessary hadronisation/decays. 
In  \PYTHIA, partons entering the hard process  are not collinear with the hadron beams. 
In the case of the charged-current DY processes, \PYTHIA\ provides,
in its event record,  the momenta of the two effective on-mass-shell quarks producing
$W$-bosons. 
A vectorial sum of these momenta gives a momentum of an appropriate $W$-boson.
Such a $W$-boson, in contrast to the case of a ME-type MC generator, carries 
the  transverse momentum, being  a vectorial sum of the quarks transverse momenta. This
has to be taken into account in constructing the hard-process event kinematics.
Let us call the kinematics in which the incoming partons are not
parallel to the beams (as a result of the aforementioned effects) 
the {\em PS kinematics}, and a corresponding MC generator -- the PS generator. 

While interfacing the ME-type generator with the PS generator one has to take 
care of the appropriate matching of the {\em ME kinematics}
with the {\em PS kinematics}. This is particularly important for processes
in which particles with non-zero spin, e.g. $W/Z$-bosons, are produced as intermediate states.  
In the following,  we describe in detail how such
a matching is performed in the \WINHAC\ interface to \PYTHIA~{\sf 6.4}.
Then we discuss possible pitfalls of the kinematical matching of
these two types of MC generators.

\vspace{2mm}
In the \WINHAC\ interface to \PYTHIA\ the final hard-process event kinematics
is constructed through the following steps:
\begin{enumerate}
\item
The {\em ME kinematics} of a given MC event is generated in \WINHAC --
the four-momenta of:  the incoming quarks, the intermediate $W$-boson,  
the final-state leptons, and the radiative photons are constructed in
the frame in which quarks are collinear with the hadron beams, where
the $+z$ axis is the direction of  one of the beams at the collision point.
\item
All the above four-momenta are then Lorentz-boosted along the $W$-boson
direction to the $W$-boson rest frame, 
which is also the centre-of-mass frame of the incoming quarks.
In this frame quarks are still aligned along the $z$ axis.
\item
The {PS kinematics} is generated by \PYTHIA, in which the effective 
on-mass-shell quarks producing the $W$-boson are non-collinear with the beams
as a results of the initial-state QCD/QED parton shower%
\footnote{Actually, for technical reasons \PYTHIA\ performs the 
so-called backward QCD evolution. This aspect is not important for our discussion.
Here we are concerned mainly with the {\em PS kinematics} in which partons 
entering the hard process are not parallel to the hadron beams -- this
can be a result of any type of a parton-shower algorithm in which transverse
degrees of freedom are not neglected (integrated out).}. 
Their four-momenta are given in the LAB
frame with the $+z$ axis along one of the hadron beams.
\item
The above \PYTHIA-quarks four-momenta are Lorentz-boosted along the sum
of their momenta to their centre-of-mass frame, 
which is also the $W$-boson rest frame%
\footnote{Instead of a single parallel boost one might use a combination of
two boost: along $p_z^W$ and along $p_T^W$.
We have checked numerically that both methods are fully equivalent.}. 
Contrary to the \WINHAC\ quarks in point 2, their momenta, although 
back-to-back, are not aligned with the $z$ axis in this frame. 
Instead, their direction 
is rotated with respect to the $z$ axis by the  polar angle $\theta_q$, 
and with respect to the $x$ axis by the  azimuthal angle $\phi_q$.
\item
From the \PYTHIA-quarks momenta, specified  in the above frame,  we calculate the
angles $(\theta_q,\phi_q)$ and then we perform rotations of all the 
\WINHAC\ momenta specified in the $W$-rest frame (point 2) using  these rotation angles. 
After such rotations the \WINHAC\ quarks are aligned with the \PYTHIA\ quarks.
\item
Finally, the whole \WINHAC\ event is Lorentz-boosted from the above frame 
to the LAB frame along the sum of the original \PYTHIA-quarks momenta 
(the boost is opposite to the one in point 4).
\end{enumerate}
In our opinion all the above steps are needed for a proper matching of
the {\em ME kinematics} with the {\em PS kinematics} in any interface between 
the ME-type MC event generator and the PS generator.  
This is particularly important for the production of the 
$W$-bosons which are spin-$1$ particles with $V-A$ couplings to fermions.
In such a case respecting all the spin correlations in the above matching
is obligatory.

The above kinematical matching relies on the correct  \PYTHIA\ generation 
of the incoming ``effective'' quarks momenta. Their spacial orientation is  crucial
for the spin correlations which, in turn, influence the angular distributions of the $W$-decay 
leptons and, as a consequence,  their  $p_T^l$ distribution.

At this point one may ask if such effective on-shell quarks 
make sense at all. It has been known for some time that a cross
section corresponding to the real NLO QCD emission can be expressed as 
a linear combination of the LO matrix elements  
\cite{Kleiss:1986re,Seymour:1994we}. 
The latter can be calculated using some effective incoming on-shell partons
four-momenta, e.g. with the help the spin amplitudes. 
However, care must be taken while constructing these effective
four-momenta for each LO matrix element individually in order not to spoil
spin correlations. Coefficients of these matrix elements can be expressed
as functions of variables related to the radiated partons (e.g. their
momenta fractions and polar angles in an appropriate frame) and they are
generally different for each LO matrix element. In a Monte Carlo approach,
computations of the NLO cross section can be done with the use of the so-called
branching algorithm, where in each branch a single LO matrix element is 
evaluated and a particular branch is picked up with a probability proportional
to the coefficient of this matrix element. What is also interesting,
flavours of these effective partons depend only on the respective LO 
process and can be different than flavours of partons initiating the NLO 
process.
Recently, such a  method has been adapted to implement the Drell--Yan processes
in the \POWHEG\ generator \cite{Hamilton:2008pd}.

For the DY processes the differential cross section corresponding to the
NLO real-parton radiation can be expressed as the following combination
of the LO matrix elements:
\begin{equation}
d\sigma_{NLO}^R = \sum_i C_i \,
|{\cal M}_{q\bar{q}'}^i(\tilde{p}_q,\tilde{p}_{\bar{q}'})|^2,
\label{eq:dsig-NLO}
\end{equation}  
where $C_i$ are the coefficients depending on the radiated parton variables,
e.g. $C_i = C_i(x,\cos\theta)$, where $x$ and $\theta$ are the momentum
fraction and the polar angle of this parton. The effective four-momenta
$\tilde{p}_q$ and $\tilde{p}_{\bar{q}'}$ of the incoming on-shell quark
$q$ and antiquark $\bar{q}'$ entering the LO matrix element are 
constructed in the NLO-process CM frame in such a way that the momentum 
of the ``spectator'' of the radiation is only rescaled without changing 
its direction, while the four-momentum of the ``emitter'' is calculated 
as a difference of the electroweak boson four-momentum and the ``spectator'' 
effective four-momentum, so its direction is different from that of 
the original ``emitter'' parton. The ``emitter--spectator'' assignment to
the incoming partons is done on the Feynman-diagramatic basis, details can be
found e.g. in \cite{Seymour:1994we,Hamilton:2008pd}. 

In order to compute the appropriate LO matrix elements and to match the LO kinematics with
the NLO one, appropriate Lorentz transformations should be performed
for the effective on-shell quarks four-momenta.
Actually, they are analogous to the ones described above for the
\WINHAC--\PYTHIA\ matching, see \cite{Hamilton:2008pd}.
Based on this analogy we believe that our procedure for the kinematical matching between 
\WINHAC\ and \PYTHIA\ should be correct, at least up to the NLO. Of course,
the parton shower provides only a leading-log (LL) type approximation of the
NLO QCD corrections, but in \PYTHIA\ the exact NLO matrix elements for 
real-parton emission can be taken into account through appropriate correcting 
weights. 
If this is done, the predictions of \PYTHIA\ for the leptonic distributions
in the DY processes should be exact at the NLO for the hard process, except for
the normalisation. The latter will not be correct because \PYTHIA\ does not 
include the NLO virtual corrections -- this could  be easily fixed by applying 
the NLO $K$-factor. 
For the lepton charge asymmetries this $K$-factor is even not needed because it 
cancels out between numerators and denominators. 
In the discussed  \WINHAC\ to \PYTHIA\ interface these NLO matrix element corrections 
are included by default.
However, the \PYTHIA\ predictions will not be exactly the same as from
the fixed-order NLO calculations, because \PYTHIA\ generates also
the higher-order LL-type QCD corrections through parton showers.
They will lead to additional distortions of the leptonic distributions, 
in particular that of $p_T^l$, but should not change them drastically.

\section{Numerical results and discussion}
\label{sec:results}

An observable which is most sensitive to details of the kinematical matching
between the ME-type MC generator and the PS generator in the charged-current 
DY process  is the final-state charged
lepton transverse momentum $p_T^l$. This is because its distribution is
a strongly varying function with a sharp Jacobian peak. Its shape is considerably
affected by the non-zero $W$-boson transverse momentum $p_T^W$, see e.g.
\cite{Fayette:2009zi}. 
Moreover, since $W$ is a vector boson and its coupling to fermions are 
of the $V-A$ type, the angular distributions of its decay leptons in the $W$-rest 
frame are highly  asymmetric:
\begin{equation}
\frac{d\sigma}{d\cos\hat{\theta}_{lq}}\; \propto\; 
\left(1 - Q_W\cos{\hat\theta}_{lq}\right)^2,
\label{eq:thetalq}
\end{equation}
where $\hat{\theta}_{lq}$ is, in the limit of massless quarks,  the angle between the outgoing charged lepton
and incoming quark directions, and $Q_W=\pm 1$ is the $W$-boson electric charge
in the units of the positron charge. 
Because of that, $p_T^l$ depends strongly not only on $p_T^W$ but also on 
the individual momenta of the quark and antiquark. 

\begin{figure}[!ht] 
  \begin{center}
    \includegraphics[width=1.04\tw]{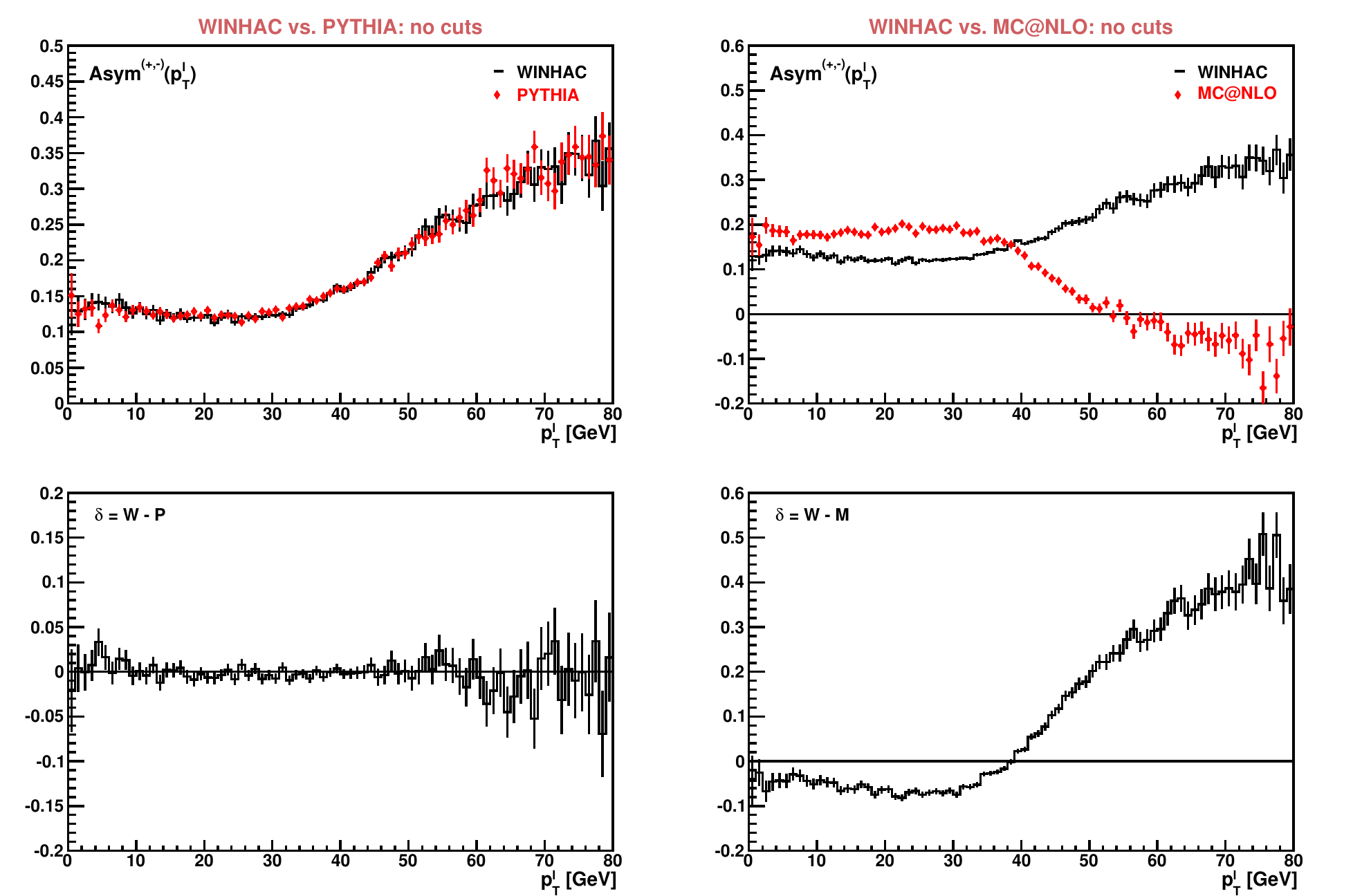}
    \caption[]
       {\figtxt{The comparisons of the $p_T^l$ charge asymmetries 
                from \WINHAC\ and \PYTHIA\ {\sf 6.4} (left plots), 
                and from \WINHAC\ and \MC@NLO\ (right plots);  
                lower plots show the 
                differences between the programs.}}
       \label{ptlas:WPM}
  \end{center}
\end{figure}

In our numerical comparisons we have used the following MC programs:
\WINHAC~{\sf 1.35} \cite{WINHAC:MC}, \PYTHIA~{\sf 6.401} \cite{PYTHIA:MC},
\MC@NLO~{4.03} \cite{MCatNLO:MC} and {\sf MCFM~5.8} \cite{MCFM:MC}.

The lepton charge asymmetry observables are  used to scrutinize the differences between the $\Wp$ and $\Wm$ mediated 
processes in these generators. For a given kinematical variable $a$, the charge asymmetry $\mathrm{Asym}^{(+,-)}(a)$ 
is defined as: 
\begin{equation}
    \Asym{a} \; =\; 
    \frac
        { \flatDfDx{\sigma^\plus}{a} - \flatDfDx{\sigma^\minus}{a} }
        { \flatDfDx{\sigma^\plus}{a} + \flatDfDx{\sigma^\minus}{a} },
        \label{eq_def_charge_asym}
\end{equation}
where $\plus$ and $\minus$ refer to the electric charge of the $W$~boson (or the final-state
charged lepton)  and $\flatDfDx{\sigma^\pm}{a}$ is the differential cross section of an observable $a$.

The asymmetry distributions  have been obtained for the proton--proton collisions at 
$\sqrt{s} = 14\,$TeV using the CTEQ~6.1 PDF set \cite{CTEQ6.1:2003}
and the particles properties from the PDG~2011 publication \cite{PDG:WWW}, 
for the following two cases: (1) without any kinematical restrictions for the outgoing lepton and (2) with the kinematical cuts
\begin{equation}
p_T^l > 20\,\textrm{GeV},\qquad
|\eta_l| < 2.5, \qquad
E_T^{\rm miss} > 25\,\textrm{GeV}.
\label{eq:cuts}
\end{equation}
 
\begin{figure}[!ht] 
  \begin{center}
    \includegraphics[width=1.04\tw]{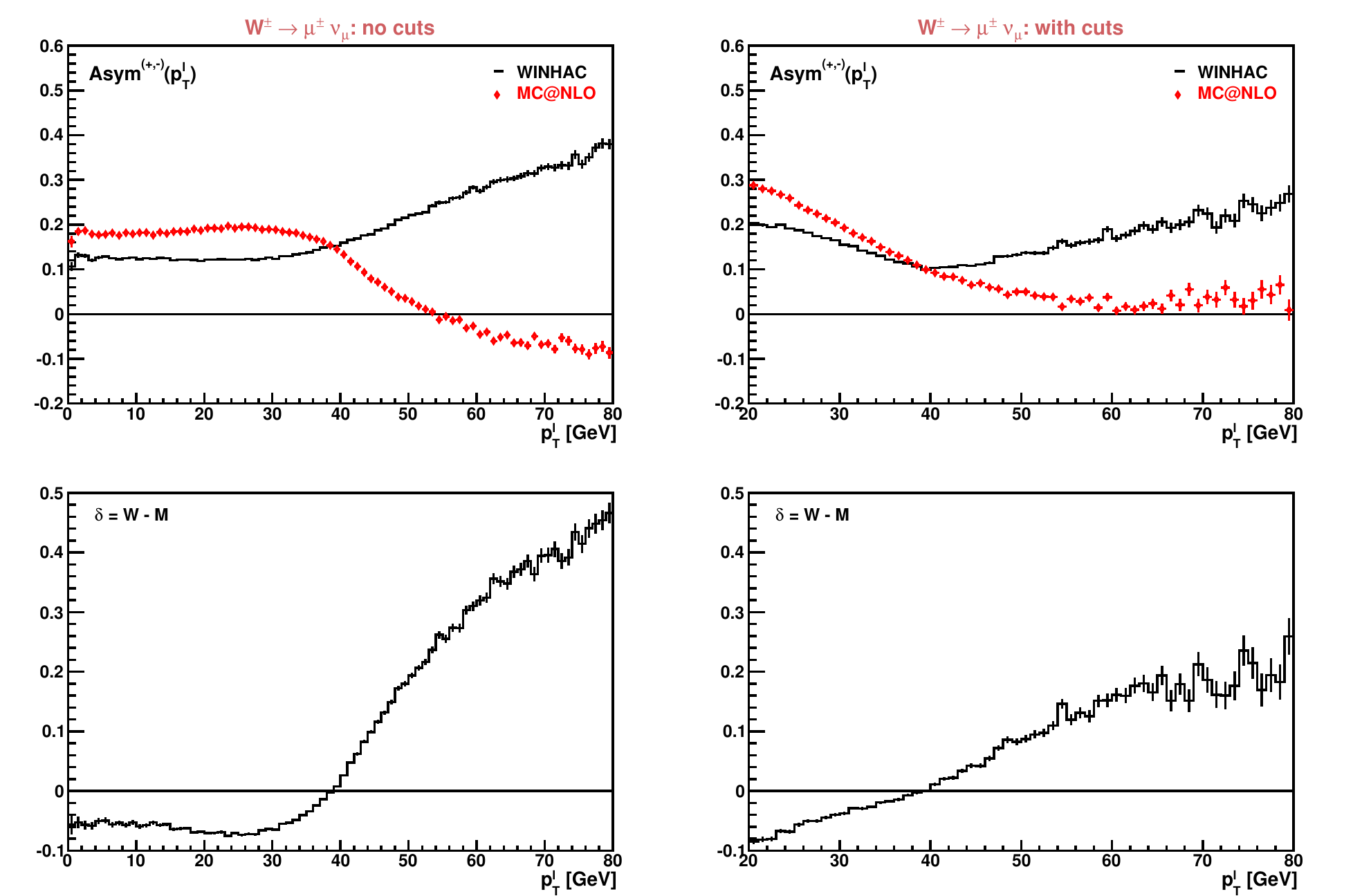}
    \caption[]
       {\figtxt{The $p_T^l$ charge asymmetries 
                from \WINHAC\ and \MC@NLO\ without cuts (left plots) 
                and with the typical ATLAS and CMS cuts (right plots);  
                lower plots show the
                differences between the programs.}}
       \label{ptlas:WM}
  \end{center}
\end{figure}

\begin{figure}[!ht] 
  \begin{center}
    \includegraphics[width=1.04\tw]{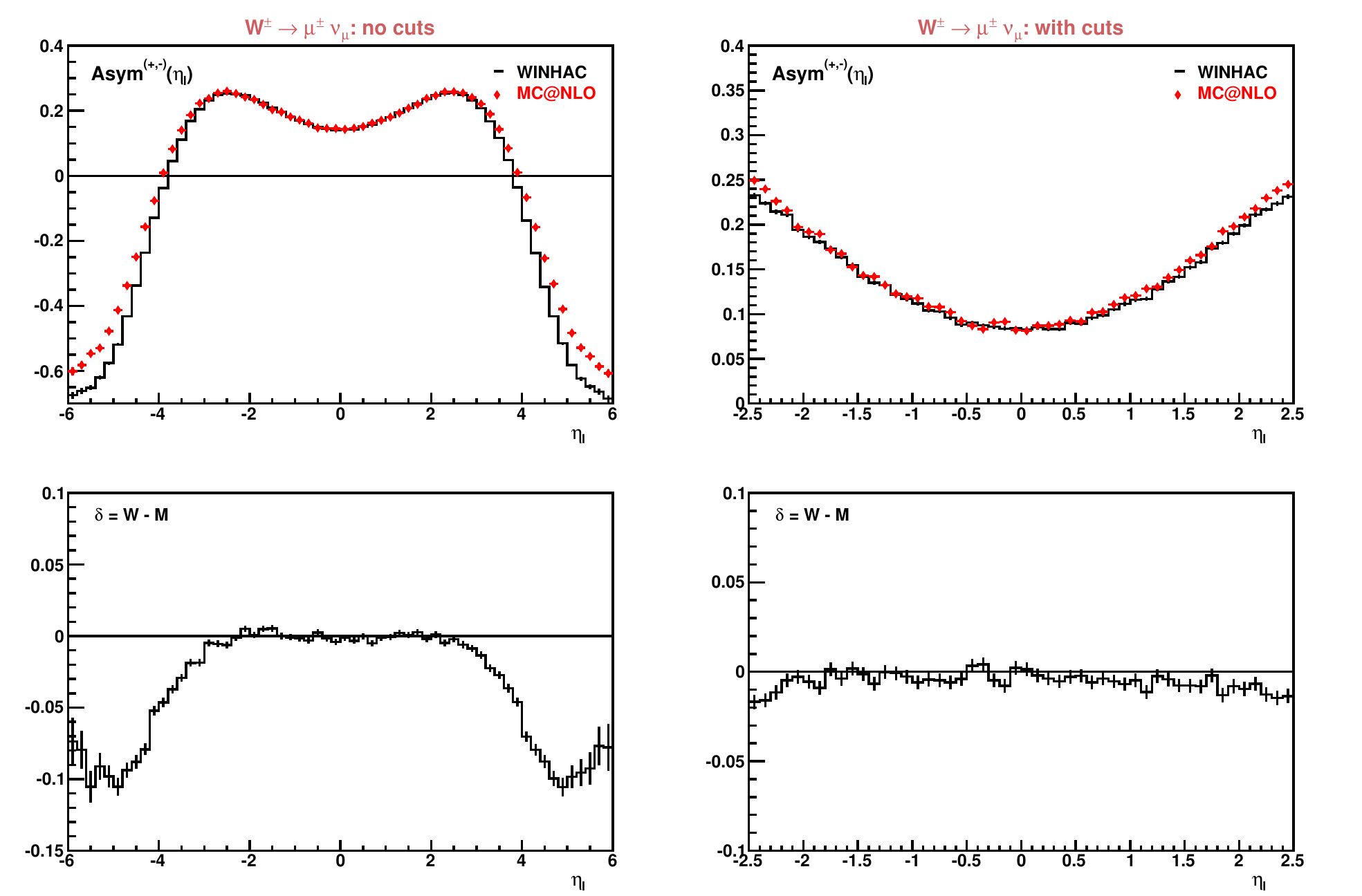}
    \caption[]
       {\figtxt{The $\eta_l$ charge asymmetries 
                from \WINHAC\ and \MC@NLO\ without cuts (left plots) 
                and with the typical ATLAS and CMS cuts (right plots);  
                lower plots show the
                differences between the programs.}}
       \label{etalas:WM}
  \end{center}
\end{figure}

In Fig.~\ref{ptlas:WPM} we show the charge asymmetry distributions as a function of  $p_T^l$ 
for electrons obtained from \WINHAC\ interfaced with \PYTHIA. 
The kinematical matching described the previous section was applied
with the effective on-shell quarks four-momenta as provided by \PYTHIA. 
These distributions
are compared with the ones coming directly from \PYTHIA\ {\sf 6.4} (left plots)
and the ones obtained from \MC@NLO\ (right plots). Lower plots show the 
differences of the distributions presented in the upper plots. 
A very good agreement between \WINHAC\ and \PYTHIA\ shows that all the technical aspects of the 
kinematical matching in their interface were done correctly.
However, we see that the \WINHAC\ and \PYTHIA\ results differ considerably from
the \MC@NLO\ results, in particular we observe 
the opposite behaviour of the $p_T^l$ asymmetry above the Jacobian peak
($\gtrsim 40\,$GeV). The region around the Jacobian 
peak is crucial for the $W$ mass measurements at the LHC,
see e.g.\ Refs.~\cite{Fayette:2008wt,Krasny:2010vd}.
The source of this discrepancy must be  understood to hope for any  improvement  
of the present precision of the the $W$-boson mass, width and their charge asymmetries at the LHC. 

In Fig.~\ref{ptlas:WM} we present the comparisons of the $p_T^l$ 
charge asymmetry distributions
for muons without kinematical cuts and with the cuts of eq.~(\ref{eq:cuts}) 
between \WINHAC\ (with the same kinematical matching 
as in Fig.~\ref{ptlas:WPM}) and \MC@NLO. 
Similar discrepancies as for the electrons are observed for the fully inclusive distributions. 
In the presence of cuts the asymmetry distribution changes its shape 
and the differences between the two programs are smaller but 
still unacceptable. 

We have also compared the charge asymmetry distributions as a function of:
$p_T^W,\,y_W$ and $\eta_l$, where:
\begin{eqnarray}
p_T^W &=& \sqrt{(p_x^W)^2 + (p_y^W)^2}\,,\\
y_W  &=& \frac{1}{2}\,\ln\left(\frac{E^W+p_z^W}{E^W-p_z^W}\right)\,,\\
  \etal &=& -\ln\left(\tan(\thetal/2)\right)\,.
\label{eq_obs_W}
\end{eqnarray}
and found a good agreement between \WINHAC\ and \MC@NLO. 
In Fig.~\ref{etalas:WM} we present charge asymmetries for $\eta_l$.
Except for the large values of $\eta_l$, i.e. except for the region which is beyond 
the measurement domain of the ATLAS and CMS experiments, 
the agreement between the two programs is good. 
The large discrepancies between the two programs in the restricted 
phase-space,  as specified by  eq.~(\ref{eq:cuts}), are thus important  only
for the charge asymmetries of the transverse lepton momentum distributions. 
 
One may argue that these discrepancies result from differences in the shape of 
the $p_T^W$ distributions between \WINHAC\ and \MC@NLO. Indeed, we have found that
the $p_T^W$ distributions differ for \WINHAC\ and \MC@NLO, but mainly at 
low $p_T^W$ ($\lesssim 6\,$GeV), while for higher values their ratio is flat.
In the \PYTHIA\ PS algorithm used by \WINHAC\ the $p_T^W$ distribution at low 
values is affected mainly but the so-called intrinsic partonic  $k_T$ which
is generated from a  Gaussian distribution with an adjustable width.   
We have used this dependence and generated the samples of events with amplified 
differences of the $p_T^W$ distributions between the above two generators in the range which is  well beyond 
the present measurement uncertainties. We have observed that  the corresponding 
charge asymmetries of the $p_T^l$ distributions remained hardly changed. We have also compared 
these asymmetries for $p_T^W> 6\,$GeV, where the ratio of the $p_T^W$ 
distributions from \WINHAC\ and \MC@NLO\ is flat, and found similar results.
Finally, have checked that, in spite of differences in the absolute $p_T^W$ 
distributions, their charge asymmetries agree very well between the two
programs.
This proofs that the differences in $p_T^W$ do not explain the large 
discrepancies in the $p_T^l$ asymmetries between \WINHAC\ and \MC@NLO.
Thus, the latter must be attributed to the differences in the effective polarisation of the 
\WINHAC\ and \MC@NLO\ $W$-bosons. 

\begin{figure}[!ht] 
  \begin{center}
    \includegraphics[width=1.04\tw]{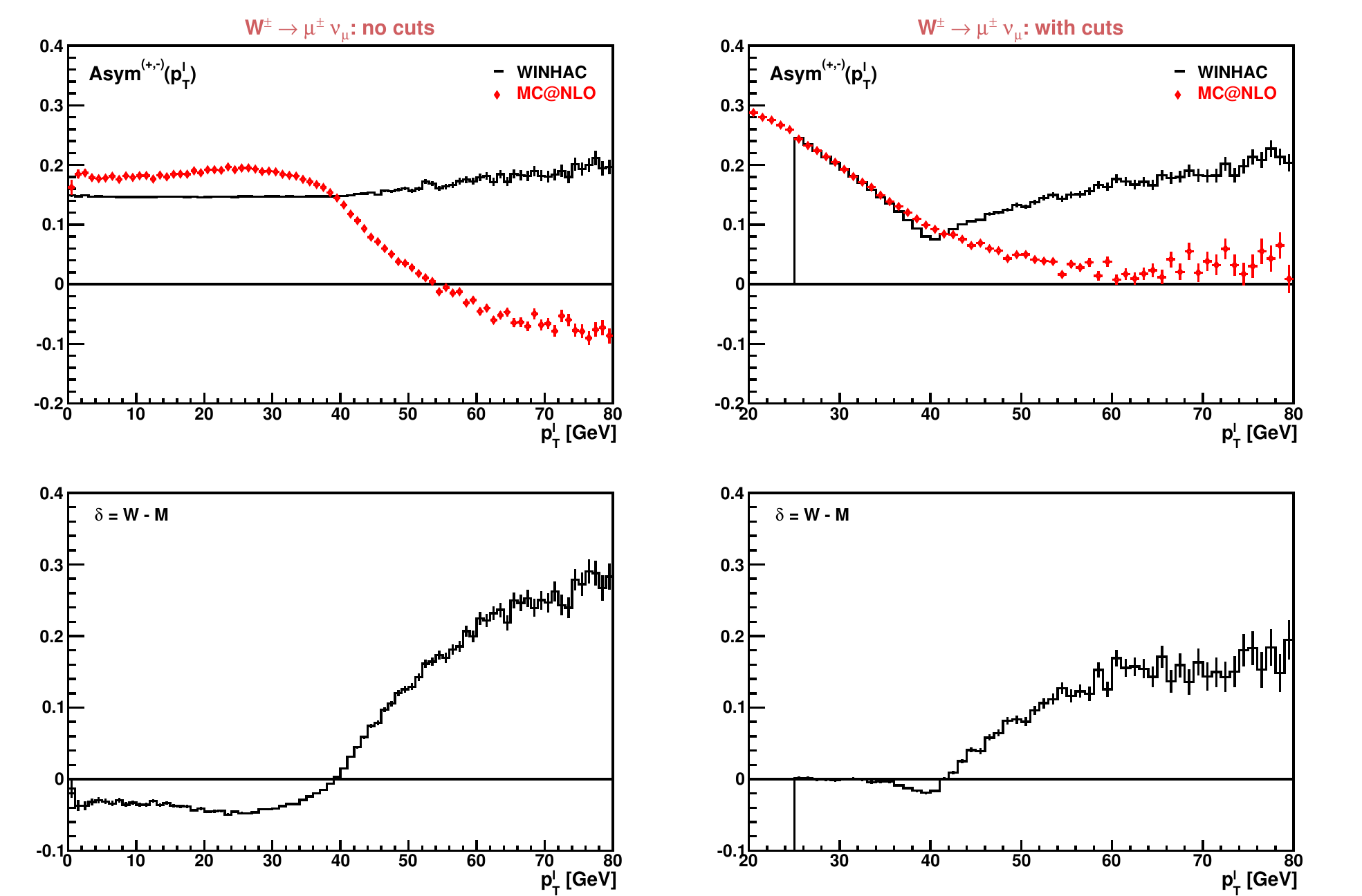}
    \caption[]
       {\figtxt{The $p_T^l$ charge asymmetries 
                from \WINHAC\ with $p_T^W=0$ and \MC@NLO, without cuts 
                (left plots) 
                and with the typical ATLAS and CMS cuts (right plots);  
                lower plots show the
                differences between the programs.}}
       \label{ptlas-pTW0:WM}
  \end{center}
\end{figure}

Can we  find simple physical arguments to explain the shape of 
the $p_T^l$ charge asymmetry distribution? 
Can we say which program is right and which is wrong? 
In order to try to answer the above two questions, we first produced the \WINHAC\ distributions for  the case
of $p_T^W =0$, i.e. without the \PYTHIA\ parton shower in \WINHAC\
(using purely beam-collinear quarks from the standard PDFs).
These  distributions are  shown in Fig.~\ref{ptlas-pTW0:WM}
and compared with the previous \MC@NLO\ result.
As one can expect, for the \WINHAC\ $p_T^W =0$ case the charge asymmetry distribution is 
flat below the Jacobian peak position
($\approx 40\,$GeV), and then rises very slowly with increasing $p_T^l$.
Its average value below the peak position reflects the difference between the total 
cross sections for positively and negatively charged DY processes 
which is driven by the effective excess of the $u$-quarks with respect 
to the $d$-quarks producing the $W$-bosons at the LHC. 
For the $p_T^l$ values above the Jacobian peak $W$-bosons  must be off-shell if  $p_T^W =0$.
Since higher invariant mass prefers harder quarks and since
$u$ is on average harder than $d$, the relative number of produced  $W^+$ rises with 
respect to  $W^-$. In the following,  this effect will be called the isospin effect.

In the presence of the kinematical cuts we see
a good agreement between the two programs below the Jacobian peak,
and the discrepancy begins above this peak but is smaller than without
cuts and also than when \PYTHIA\ is used.
The sharp cut at $p_T^l=25\,$GeV for the \WINHAC\ results comes from the 
cut on $E_T^{\rm miss}$ (in the case of $p_T^W=0$ they are equivalent). 
It is rather striking  that below the Jacobian peak the charge asymmetry of the $p_T^l$ 
distribution is  generated
mainly by the cut on the lepton pseudorapidity,  $\eta_l$,  and is hardly sensitive to the  
$W$-bosons transverse momentum spectrum. 

What is the reason for the observed shape of the distribution for $p_T^W=0$? 
In the $W$-rest frame the events with high $p_T^l$
correspond to $\eta_l \approx 0$, while the ones with low $p_T^l$
to large positive and large negative $\eta_l$. If we take the $+z$ axis
along the quark momentum, then the $W$-bosons will have preferably 
positive rapidity in such a frame, since the quarks are on average harder than
the antiquarks. Thus, when we perform a boost to the LAB frame in the presence 
of symmetric cuts on $\eta_l$, the events with negative $\eta_l$ will migrate 
in while the ones with positive $\eta_l$ will migrate out of the selected kinematical region
(the $W$-boson rapidity just adds to the lepton pseudorapdity).
Since for $W^-$ charged leptons are emitted preferably along its direction,
then more events with low $p_T^l$ will move out than move in, while
for $W^+$ it will be opposite. This is why we observe the decrease of the 
asymmetry with increasing $p_T^l$
up to the value close to the Jacobian peak position.  Close to the peak position majority of leptons must  have 
 $\eta_l \approx 0$ in the $W$-rest frame and the migration mechanism discussed above can be neglected.
 For such $p_T^l$ values and above the discussed earlier mechanism related to the relative hardness 
 of the distributions of the $u$ and $d$ quarks takes over and the asymmetry rises.

Having understood the influence of the migration and the quark-isospin effects on the 
charge asymmetry distribution,  
let us try to answer our main question: do we understand  the $p_T^l$
charge asymmetry when $p_T^W>0$? 

As discussed before, the shape of the distribution below the Jacobian peak position 
is determined  by the migration mechanism and is hardy dependent on the 
underlying distribution of $p_T^W$. Therefore, in this region,  our previous analysis holds. 
We thus concentrate  on the region of the large  $p_T^l$ (above 
the position of the Jacobian peak). The main difference here with respect to the 
$p_T^W=0$ case is that in addition to the isospin effect another effect come into play and become dominant: 
the effect of hard QCD radiation which influences  the  effective polarisation of the 
$W$-bosons. In the discussion presented below  the $W$-polarisation is specified  in the reference frame 
in which the spin quantisation axis is parallel to the direction of the $W$-boson. 

It has been shown recently that for the processes
of $W+ jets$ production at the LHC left-handedly polarized $W$s dominate 
over the right-handedly polarized $W$s \cite{Bern:2011ie}. 
For the left-handed $W$s the charged leptons are emitted
preferably in the $W^-$ direction and opposite to the $W^+$ direction (and {\em vice versa} for the right-handed $W$s).
Therefore, the  non-zero $p_T^W$ increases, on the average,  the transverse momentum
of the negatively charged lepton and decreases  it for the positively charged one. This is what we observe
in the left plot of Fig.~\ref{ptlas-pTW0:WM} where for \MC@NLO\  
the asymmetry decreases for high $p_T^l$. There is, of course, some contribution from
longitudinally polarized $W$s, but it never dominates \cite{Bern:2011ie} 
and, what is more important, charged lepton angular distributions 
are in this case identical for $W^+$ and $W^-$. Moreover,  
the isospin effect, which could potentially counterbalance such a decrease, is 
sizeably smaller in magnitude due to a steeply falling Breit--Wigner distribution.
Therefore, the \MC@NLO\ results do have a rather convincing explanation 
of the $p_T^l$ charge asymmetry behaviour
while the \PYTHIA\ results do not. 

\begin{figure}[!ht] 
  \begin{center}
    \includegraphics[width=1.04\tw]{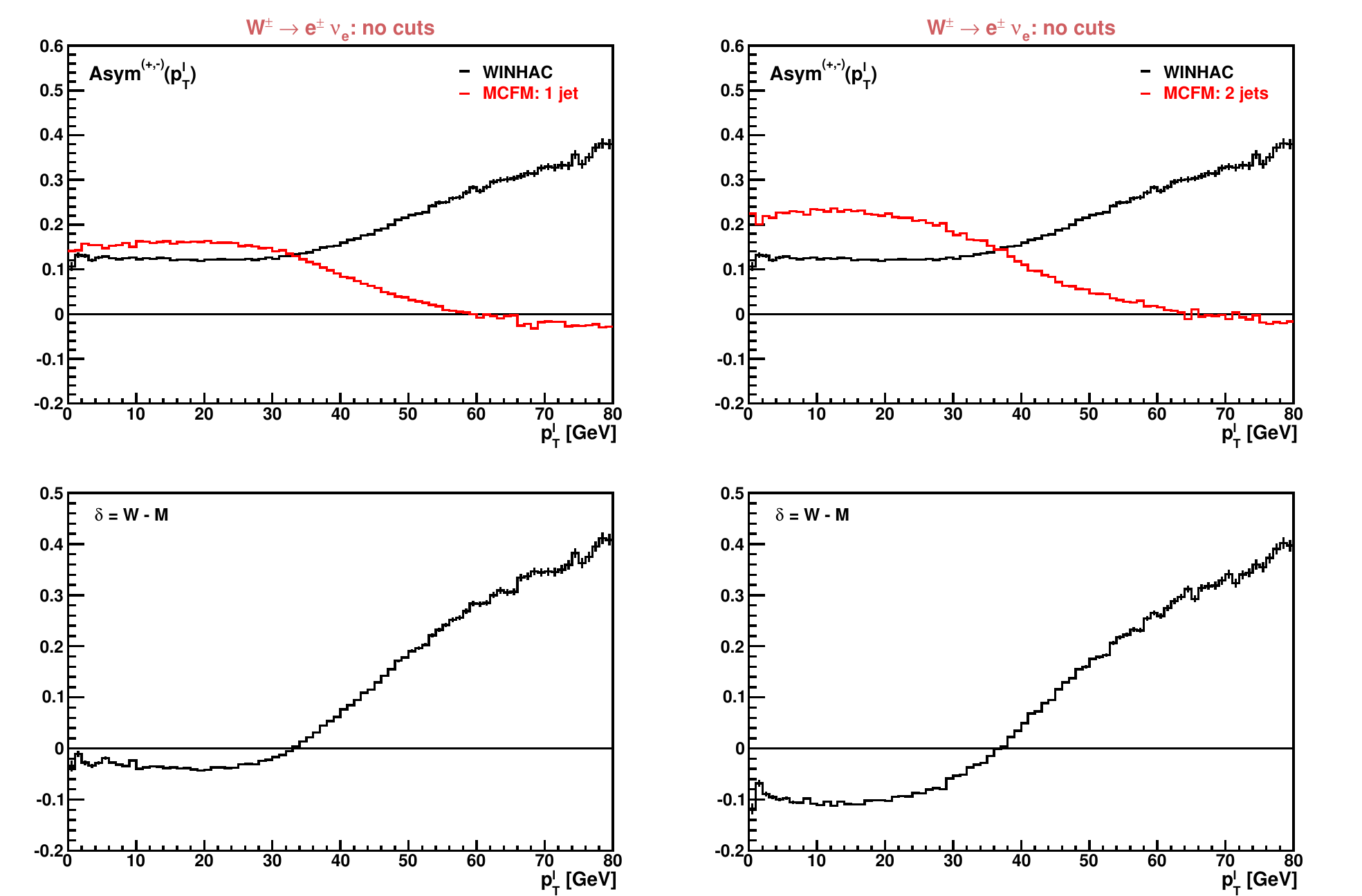}
    \caption[]
       {\figtxt{The $p_T^l$ charge asymmetries 
                from \WINHAC\  
                and {\sf MCFM} for $W+1\:jet$ (left plots), 
                and {\sf MCFM} for $W+2\:jets$  (right plots);  
                lower plots show the
                differences between the programs.}}
       \label{ptlas:MCFM}
  \end{center}
\end{figure}

In order to check the validity of the above reasoning we have generated events using 
the Monte Carlo program {\sf MCFM}  \cite{MCFM:MC} which calculates the fixed-order
QCD corrections to the hard process convoluted with the collinear PDFs.
The charge asymmetry distribution for $W+1\:jet$ and $W+2\:jets$ are presented in 
Fig.~\ref{ptlas:MCFM} and compared with the ones from \WINHAC\
with the standard \PYTHIA\ parton-shower matching. The $p_T^l$ asymmetries
predicted by {\sf MCFM} are close to those from \MC@NLO, which supports 
the conclusion that the \MC@NLO\ predictions
on the $p_T^l$ asymmetries are more likely to be correct than those of  \PYTHIA. 

\begin{figure}[!ht] 
  \begin{center}
    \includegraphics[width=1.04\tw]{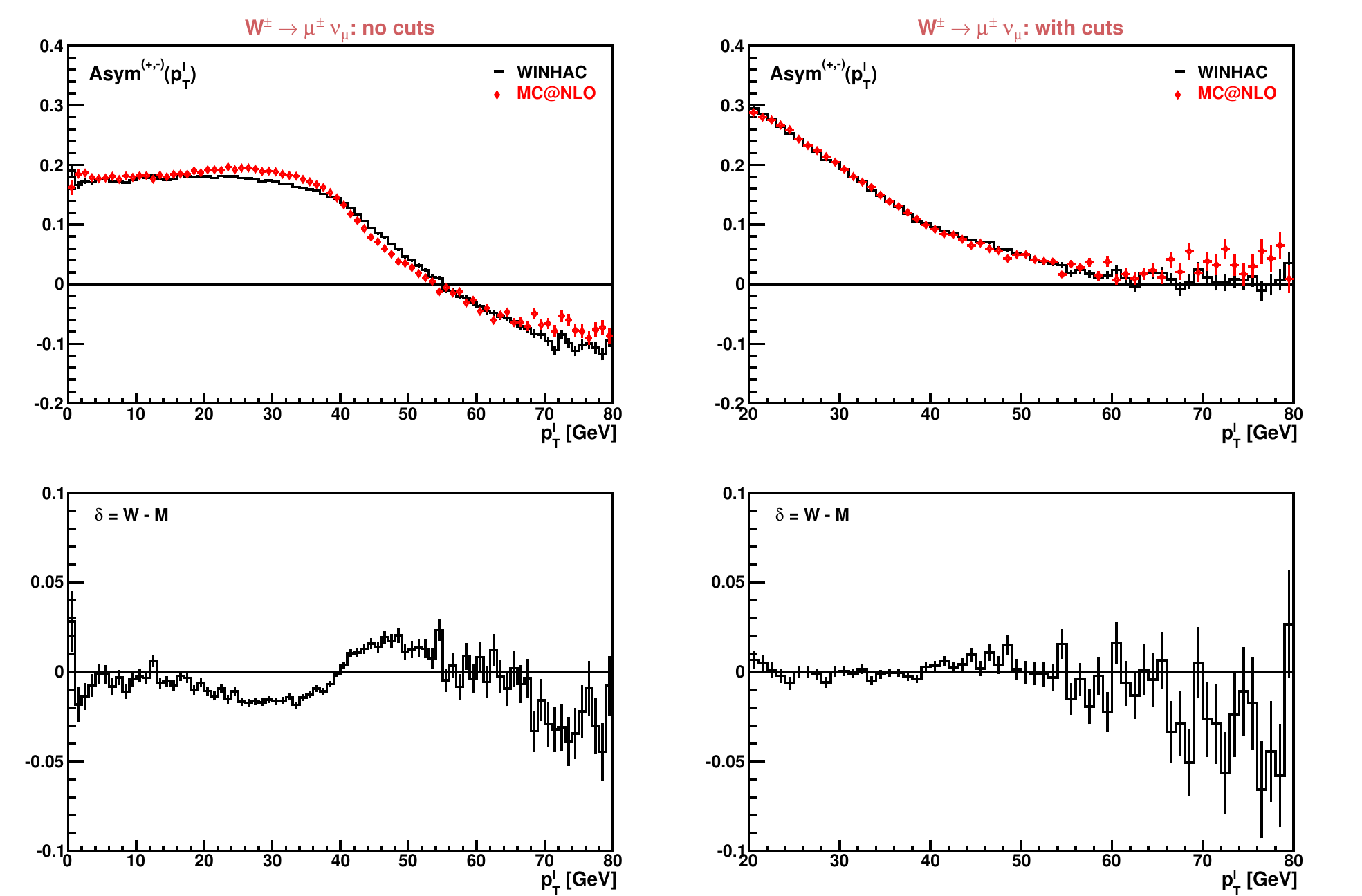}
    \caption[]
       {\figtxt{The $p_T^l$ charge asymmetries 
                from \WINHAC\ with the transverse
                momenta of the effective quarks {\em swapped} 
                and \MC@NLO, without cuts (left plots) 
                and with the typical ATLAS and CMS cuts (right plots);  
                lower plots show the
                differences between the programs.}}
       \label{ptlas:WM-kTswap_ptl}
  \end{center}
\end{figure}

\begin{figure}[!ht] 
  \begin{center}
    \includegraphics[width=1.04\tw]{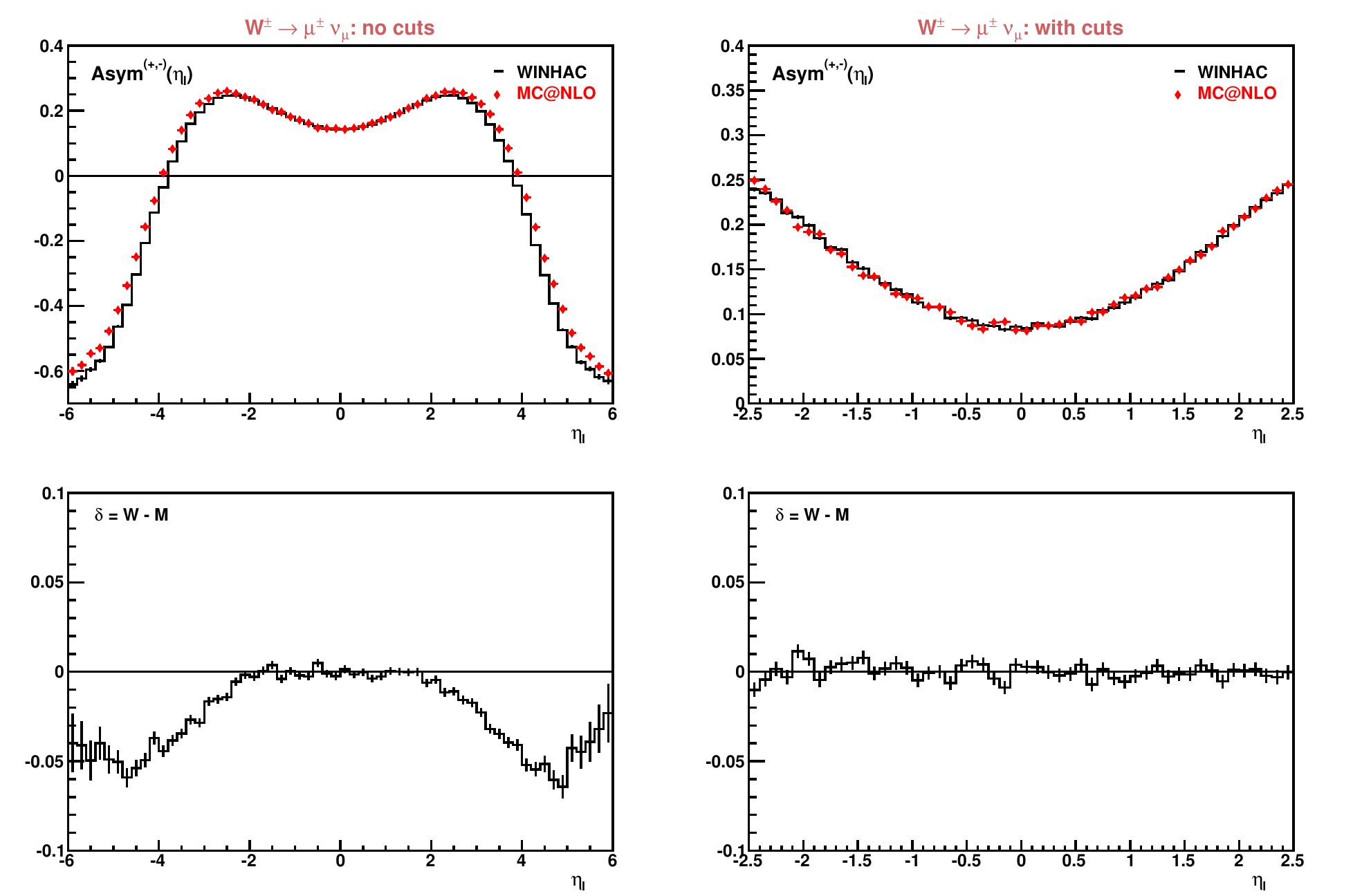}
    \caption[]
       {\figtxt{The $\eta_l$ charge asymmetries 
                from \WINHAC\ with the transverse
                momenta of the effective quarks {\em swapped} 
                and \MC@NLO, without cuts (left plots) 
                and with the typical ATLAS and CMS cuts (right plots);  
                lower plots show the
                differences between the programs.}}
       \label{ptlas:WM-kTswap_etal}
  \end{center}
\end{figure}

But can we find the simple reason why the \PYTHIA\ predictions are so grossly wrong?
From  our numerical tests and discussion presented above it becomes 
obvious that the problem must be related to 
the modelling of the effective polarisation of $W$-bosons.
In the LO approximation and for on-shell partons the $W$-polarisation  is uniquely driven by
the asymmetry in the  distributions of 
the momenta of the effective quark and antiquark
entering the DY processes  (see eq.~(\ref {eq:dsig-NLO})),
rather than by their sum which determines  $y_W$ and $p_T^W$.
Inspecting the \PYTHIA~{\sf 6.4} manual \cite{Sjostrand:2006za} we
have found that the construction of these effective on-shell partons momenta 
should agree at NLO with that of Ref.~\cite{Hamilton:2008pd}. 
\PYTHIA, of course, generates through the parton-shower more than a single 
NLO emission, however they should not change considerably (or even revert) the NLO effective 
partons momenta as such additional emissions are mainly soft and collinear. 
Therefore, the \PYTHIA\ predictions for the $p_T^l$ asymmetries should
not differ much from the NLO ones.

At this point we started searching not only for possible conceptual but also for the  technical errors
affecting the spatial orientation of the quark and antiquark momenta. 
We have made several technical checks of the \PYTHIA\ generator along this line. 
One of the checks done was to 
swap the transverse momenta of the effective on-shell
quark and antiquark. To our great surprise, once this was done on the event-by-event basis, 
we have obtained  a very good agreement with the \MC@NLO\ charge asymmetry distribution, both in the 
full phase-space and in the restricted kinematical region. The comparisons 
are shown in Fig.~\ref{ptlas:WM-kTswap_ptl} for the $p_T^l$ 
and in Fig.~\ref{ptlas:WM-kTswap_etal} for the $\eta_l$ dependence of the 
lepton charge asymmetry. 
This agreement may be accidental but  it may also suggest  that the transverse momenta are,  perhaps,  not correctly 
assigned to the effective quark and antiquark in \PYTHIA. 
Whether or not such a hypothesis is true can however be verified only by the authors 
of the \PYTHIA\ generator.

\begin{figure}[!ht] 
  \begin{center}
    \includegraphics[width=1.04\tw]{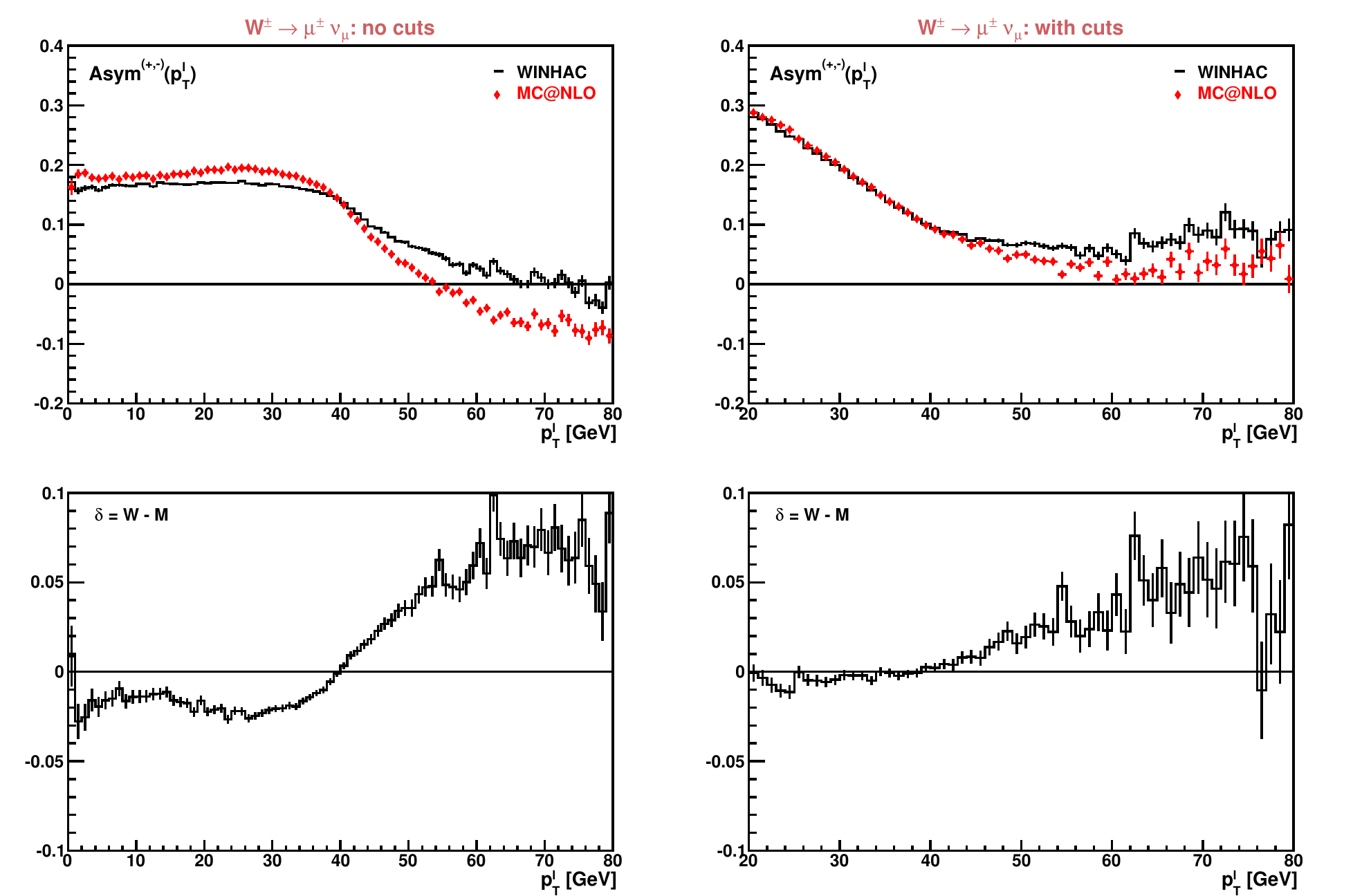}
    \caption[]
       {\figtxt{The $p_T^l$ charge asymmetries 
                from \WINHAC\ without rotations of leptons momenta
                in the $W$-rest frame 
                and \MC@NLO, without cuts (left plots) 
                and with the typical ATLAS and CMS cuts (right plots);  
                lower plots show the
                differences between the programs.}}
       \label{ptlas:WM-2boosts}
  \end{center}
\end{figure}

On the conceptual side we have  investigated the mechanism 
which drives the effective LO polarisation of the $W$ bosons  
in the DY process involving on-shell quarks. 
We have found that skipping the rotations of leptons momenta in the
$W$-boson rest frame, described in point 5 of the previous section,
gives better agreement of the \PYTHIA\ $p_T^l$ charge asymmetry
with the \MC@NLO\ one.  The results are shown in 
Fig.~\ref{ptlas:WM-2boosts}. The agreement with \MC@NLO\ is slightly
worse than in Fig.~\ref{ptlas:WM-kTswap_ptl}, however much better than in 
Fig.~\ref{ptlas:WM}. 
Note that skipping these rotations is equivalent to retaining the PS-initial (parton-shower
unaffected) effective on-shell quark helicities rather than those corresponding 
the PS-final ones (following the parton-shower). 

We have implemented the above two  options in the new version
of \WINHAC~\cite{WINHAC:MC}. 
These versions cannot replace the future state-of-the-art NLO programs 
with the NLO PS and the full set of EW radiative corrections. However, 
as long as such programs are not available, they may be of use in 
the initial phase of the measurement of the lepton charge asymmetries at the LHC.
First of all, they can be of help  in the unfolding of the measured charge lepton asymmetries
in the experimental procedures  where the precision of the EW and the real photon radiative 
correction matters. 
More importantly,  the above versions, providing the simplified LO picture of the effective polarisation 
of $W$-bosons at the LHC,  may help in designing the new polarisation-dependent observables  
for the studies of the electroweak symmetry breaking (EWSB) mechanism.

\section{Summary}

In this paper we have discussed the generic problem of kinematical matching of 
a  parton shower generator with a matrix element generator for the Drell-Yan  processes 
involving spin 1 intermediate particles. We have argued that the Les Houches Accord must  be extended 
to take into account the spin correlations at all the stages of the event generation.
We have described in detail our  kinematical matching 
procedure which is used in the interface of our \WINHAC\ generator with 
the \PYTHIA\ {\sf 6.4} generator. We have demonstrated that the momentum vectors  of the on-shell 
quark and antiquark, carrying in the LO approximation  the full information on the 
$W$-boson polarisation,  must be well defined. Any error in 
directions of these vectors has dramatic consequences for 
the $p_T^l$ dependence of the charge asymmetries 
at the LHC. In particular, using the transverse momenta of the effective
quarks provided in \PYTHIA\ {\sf 6.4} leads to
completely different behaviour of the above asymmetry than predicted
by the NLO (and beyond) calculations, e.g.\ \MC@NLO\ and {\sf MCFM}.
We have found that simple swapping of the effective quark and antiquark 
transverse momenta in \PYTHIA, or skipping the rotation of 
the outgoing  lepton momenta,   results in the $p_T^l$ charged asymmetries
that match the NLO predictions of \MC@NLO. We have implemented
the corresponding matching schemes in the new version of the \WINHAC\ generator.

The issue of the proper matching between the matrix element calculations
and the parton-shower generators respecting the spin correlations
is important not only for the charged-current Drell--Yan processes
but also for any process of production and decay of non-zero spin particles 
at the LHC. It needs to be readdressed in the more general context of matching 
the NLO matrix elements with the NLO parton shower such that a handle 
is given to the experimentalists to control the relative contributions of all the 
spin density matrix elements of the decaying particles,  thus allowing for 
an experimental verification of the implemented Monte Carlo mechanism which drives 
the polarisation of non-zero spin particles at the LHC.

\vspace{5mm}
\noindent
{\large\bf Acknowledgements}
\vspace{3mm}

\noindent
We would like to thank S.\ Jadach for useful discussions.

\vspace{5mm}
\noindent


\end{document}